\documentclass[3p,times,twocolumn]{elsarticle}

\usepackage{ecrc}

\volume{00}

\firstpage{1}

\journalname{Nuclear and Particle Physics Proceedings}

\runauth{}

\jid{nppp}

\jnltitlelogo{Nuclear and Particle Physics Proceedings}

\usepackage{amssymb}

\usepackage[figuresright]{rotating}

\newcommand{\be}{\,\begin{equation}}
\newcommand{\ee}{\,\end{equation}}

\begin{document}

\begin{frontmatter}



\dochead{}

\title{Non-linear Cosmic Ray Propagation}


\author{P. Blasi}

\address{INAF/Osservatorio Astrofisico di Arcetri, Largo E. Fermi, 5 - 50125 Firenze, Italy \\
Gran Sasso Science Institute, Viale F. Crispi, 7 - 67100 L'Aquila, Italy}

\begin{abstract}
The description of the transport of cosmic rays in magnetized media is central to both acceleration and propagation of these particles in our Galaxy and outside. The investigation of the process of particle acceleration, especially at shock waves, has already emphasized that non-linear effects such as self-generation of waves and dynamical reaction of cosmic rays on the background plasmas, are crucial if to achieve a physical understanding of the origin of cosmic rays. Here we discuss how similar non-linear effects on Galactic scales may affect the propagation of cosmic rays, not only through the excitation of plasma waves important for particle scattering, but also by inducing the motion of the interstellar medium in the direction opposite to the gravitational pull exerted by matter in the Galaxy, thereby resulting in the launching of a wind. The recent discovery of several unexpected features in cosmic ray spectra (discrepant hardening, spectral breaks in the H and He spectra, rising positron fraction and unexpectedly hard antiproton spectrum) raises the question of whether at least some of these effects may be attributed to poorly understood aspects of cosmic ray transport.
\end{abstract}


\end{frontmatter}


\section{Introduction}
\label{sec:intro}

The quest for the origin of cosmic rays (CRs) boils down to building a model of particle acceleration (in a class or classes or sources) and a model for transport that combined together may provide a consistent picture of what we observe at the Earth. In the last decade or so the community experienced the growth of considerable self-confidence in numerical approaches to CR propagation (e.g. GALPROP, DRAGON, Usine, ...) that appear to be able to provide such consistent picture. Though on one hand we should be proud of this success, on the other hand we should always question the physical ingredients that such approaches are based upon and the assumptions that have been made to make predictions consistent with observations. Far from being a criticism of such approaches, this statement is mainly aimed at making a sanity check on whether some of the assumptions adopted in these calculations are actually justified and/or whether some of them may in fact reflect some more basic physical process. For instance, basically all numerical approaches to CR propagation require breaks in the rigidity dependence of the diffusion coefficient or the injection spectrum (or both): such breaks are introduced by hands and allow to reach a picture of the origin of CRs (spectra at the Earth, secondary-to-primary ratios and integrated Galactic gamma ray emission) that most feel satisfied with. On the other hand, spectral breaks in energy dependences that are otherwise close to power laws usually suggest the existence of energy (or rather rigidity) scales in the problem. Assuming the existence of such breaks might lead us to overlook some important pieces of the CR puzzle. Another instance of this type of potentially weak links is the assumption of free escape of CRs from a halo with a given size $H$: this assumption, that mathematically translates into the boundary condition that the density of CRs at a location at distance $H$ above and below the Galactic disc vanishes, might in fact reflect the fact that there is scale height in the level of turbulence (the diffusion coefficient grows very fast for $|z|>H$) or it might reflect the transition from a diffusion dominated to an advection dominated regime, as in the case of CR induced winds. But the implications of these two scenarios in terms of, for instance, secondary-to-primary ratios, are very different. In other words, imposing this boundary condition on the transport equation, as it is routinely done, might result in overlooking an important piece of information on the transport of CRs in the Galaxy. 

There is a second, more pragmatic reason, for questioning CR transport: recent measurements have revealed several unexpected findings, from spectral breaks in the spectrum of protons and helium nuclei \cite{2011Sci...332...69A,2015PhRvL.115u1101A,2015PhRvL.114q1103A} to a rise in the positron fraction \cite{2009Natur.458..607A,2013PhRvL.110n1102A} and a rather hard spectrum of antiprotons \cite{2016PhRvL.117i1103A}. It makes sense to ask whether some or all of these findings may be explained in terms of poorly understood aspects of CR transport.

This review is organized as follows: in \S \ref{sec:standard} we briefly summarize the basic aspects of the standard model of CR transport in the Galaxy. The reader should not expect a complete review of the models on the market, but rather a summary of the essential aspects of propagation. One should keep in mind that what makes numerical codes such as GALPROP and DRAGON so useful is not their treatment of CR transport itself, which is rather standard, but the account of the many cross sections involved in the calculation of both the gamma ray emission and the production of secondaries (antiprotons, positrons and secondary nuclei). Here we will not discuss cross sections any further, although we realize that they represent a crucial ingredient in the comparison with observations. In \S \ref{sec:Gal-self} we discuss the process of streaming instability and the growth and damping of the waves that result from it. The implications of these effects in terms of Galactic CR transport will be summarized. In \S \ref{sec:Gal-wind} we discuss CR induced winds, namely CR transport under the effect of self-generated turbulence and the dynamical launching of a Galactic wind due to the force exerted by CRs on the background plasma. In \S \ref{sec:grammage-source} we discuss a rather dramatic implication of the self-generation of waves: the CR gradient near a source may be strong enough to warrant a high level of turbulence that in turn confines CRs near the source for times much longer than naively expected. This implies that CRs experience a grammage in the near source region that may become comparable with the grammage usually attributed to the propagation throughout the Galaxy. In \S \ref{sec:discuss} we discuss the global picture that arises from introducing non-linear effects in CR propagation and we speculate on their possible role in explaining some of the anomalies revealed by recent CR measurements. 

\section{A summary of standard predictions}
\label{sec:standard}

The simplest implementation of CR transport is one in which the effective transport only takes place in the $z$ direction perpendicular to the Galactic disc. The formalism can be easily generalized to 3D. The transport equation for CR nuclei (both primaries and secondaries) can be written as:
$$
-\frac{\partial}{\partial z} \left[D_{\alpha}(p) \frac{\partial f_{\alpha}}{\partial z}\right] + w\frac{\partial f_{\alpha}}{\partial z} -\frac{p}{3}\frac{\partial w}{\partial z}\frac{\partial f_{\alpha}}{\partial p}
+\frac{\mu v(p) \sigma_{\alpha}}{m}\delta(z) f_{\alpha} + 
$$
$$
\frac{1}{p^{2}} \frac{\partial}{\partial p}\left[ p^{2} \left(\frac{dp}{dt}\right)_{\alpha,ion} f_{\alpha}\right] =
$$
\be
~~~~~~~~~~~~= 2 h_d q_{0,\alpha}(p) \delta(z) +\sum_{\alpha'>\alpha} \frac{\mu\, v(p) \sigma_{\alpha'\to\alpha}}{m}\delta(z) f_{\alpha'},
\label{eq:slab}
\ee
where we defined:
\be
w(z)=(u+\bar v_{A})\Theta(z)-(u+\bar v_{A})\left[ 1 - \Theta(z) \right],
\ee
with $\Theta(z)$ the Heaviside function. Here $u$ is the velocity of a possible Galactic outflow (if none is present $u=0$). The mean Alfv\'en speed $\bar v_{A}$ is the effective Alfv\'en speed averaged over the direction of motion of the waves: if there is an equal number of waves moving in both directions then $\bar v_{A}=0$. This is the situation in which one should expect the highest degree of second order Fermi acceleration. On the other hand, if waves only move away from the disc, as it is expected to be the case if they are self-generated, then $\bar v_{A}=v_{A}=B_{0}/\sqrt{4\pi m_{p} n_{i}}\approx 15$ km/s ($B_{0}=1\mu G$ and $n_{i}=0.02~\rm cm^{-3}$ are the magnetic field strength and gas density in the halo). In this case there is no second order Fermi acceleration. 

Although many new recent results have concerned the leptonic component of CRs (positrons and electrons), here we will only focus on spectra of hadrons. It should however be kept in mind that the combined observation of the spectra of leptons and nuclei might provide important insights on CR transport (see for instance \cite{lipari}).

In Eq. (\ref{eq:slab}), $\sigma_{\alpha}$ is the spallation cross section of a nucleus of type $\alpha$, $\mu$ is a grammage parameter fixed to 2.4 mg/cm$^2$, and $q_{0,\alpha}(p)$ is the rate of injection per unit volume in the disk of the Galaxy. Namely, since $h_d$ is the half-thickness of the (assumed infinitesimal) gaseous disk, and $2\,h_d\,q_{0,\alpha}$ is the rate of injection in the disk of the Galaxy per unit surface. The total cross section for spallation and the cross sections for the individual channels of spallation of a heavier element to a lighter element ($\sigma_{\alpha'\to\alpha}$) are provided by \cite{1990PhRvC..41..566W} and \cite{2003ApJS..144..153W}. 

At high enough energies that ionization losses (fifth term in Eq. \ref{eq:slab}) are not important and advection terms, even when present, are subdominant, and assuming protons as primary nuclei, the equation above reduce to the well known equation:
\be
-\frac{\partial}{\partial z} \left[D_{p}(p) \frac{\partial f_{p}}{\partial z}\right] =
2 h_d q_{0,p}(p) \delta(z) 
\label{eq:slabprotons}
\ee
where the injection term is $2 h_d q_{0,p}(p) = \frac{N(p){\cal R}}{\pi R_{d}^{2}}$, with $N(p)$ is the spectrum produced by each source (for instance a SNR), ${\cal R}$ is the rate per unit time of occurrence of such sources in the Galaxy and $R_{d}$ is the radius of the Galactic disc. Eq. \ref{eq:slabprotons} is usually solved by imposing a free escape boundary condition, $f_{p}(|z|=H,p)=0$ and using the symmetry of the problem, $\frac{\partial f_{p}}{\partial z}|_{z=0^{-}}=-\frac{\partial f_{p}}{\partial z}|_{z=0^{+}}$. These assumptions lead to the well known solution: $f_{0,p}=\frac{N(p) {\cal R}}{2 \pi R_{d}^{2} H}\frac{H^{2}}{D(p)}$. This latter expression has a simple physical interpretation: the spectrum observed at the Earth is the product of the rate of injection over the entire volume of the Galaxy and the confinement time $H^{2}/D(p)$. For an injection $N(p)\propto p^{-\gamma}$ and a diffusion coefficient $D(p)\propto p^{\delta}$, the CR spectrum inside the disc is $f_{p}\propto p^{-\gamma-\delta}$.  

For primary nuclei the situation is similar but energy losses due to spallation may become important (see for instance \cite{2012JCAP...01..010B}), especially for heavier nuclei. This phenomenon leads to a hardening of the equilibrium spectrum of nuclei at low energies. For all nuclei, including protons, ionization losses become important at low energies. What happens in this energy region depends on a rather complex interplay between advection (when present), diffusion (in this energy region the diffusion coefficient is usually, but not always, assumed to be constant) and losses. 

If the target gas for spallation reactions is assumed to be concentrated in a infinitely thin disc, then the injection term for secondary nuclei is also $\propto \delta(z)$ and the solution for secondary nuclei is easily found. This is the standard way to demonstrate that, at sufficiently high rigidity, the secondary-to-primary ratios are $\propto X(R)$, where $X(R)\propto R^{-\delta}$ is the grammage as a function of rigidity $R$. As a rule of thumb this conclusion holds for both the $\bar p/p$ and the $B/C$ ratio (although it is good to keep in mind that the cross section for antiproton production is expected to have a stronger energy dependence \cite{2014PhRvD..90h5017D,2015JCAP...09..023G} than the spallation cross section, so as to affect the simple prediction that $\bar p/p\propto R^{-\gamma}$). 

The observed low energy behaviour of these ratios typically requires that either the diffusion coefficient has a plateau at rigidity $\lesssim 3 GV$, or that the injection spectrum flattens at low energies and substantial reacceleration occurs, or, finally, a prominent role of advection. In standard approaches to CR transport, these instances are implemented by imposing by hand breaks in either the diffusion coefficient or the injection spectrum. However, one should recall that breaks in power laws are typically a signature of the appearance of physical scales. Imposing such breaks without motivating such action may hide important physical effects, as we discuss below. 

The simple model illustrated above may be easily generalized to mildly more complex situations, such as a $z$-dependent non-separable diffusion coefficient (see for instance \cite{2002ApJ...572L.157D,2012ApJ...752L..13T}).

A residual grammage is also often introduced as a free parameter, to be added to $X(E)$. The residual grammage is usually assumed to be rigidity independent and to represent the potential contribution to secondary production inside the sources of CRs, although in some alternative scenarios of CR transport \cite{2010PhRvD..82b3009C,2014ApJ...786..124C,2016ApJ...827..119C}, most grammage is assumed to be accumulated near the source rather than during propagation through the Galaxy, and an energy independent grammage at high energy is instead assumed to be the contribution of the Galaxy. Physically, the introduction of a residual grammage implies that the secondary-to-primary ratios should become roughly constant at sufficiently high rigidity. 

\section{Galactic CR transport in self-generated waves}
\label{sec:Gal-self}

Several authors have discussed the possibility that CRs could be self-confined by the waves generated through the streaming instability that they excite in the direction of their spatial gradient (see \cite{cesarsky,wentzel} for reviews). In particular \cite{skilling} and \cite{holmes} discussed the effect of self-generation in the presence of ion-neutral damping and non-linear Landau damping (NLLD). The general conclusion in both cases is rather interesting: in the Galactic disk and its vicinities, waves are damped so fast that CR transport is ballistic, while particles are dragged at the Alfv\'en speed (confinement) only in the halo, much above or below the Galactic disc, where the neutral gas density is sufficiently low to avoid effective ion-neutral damping. However, the results of these early works were not based on a self-consistent solution of the transport equation.

In the presence of a gradient in the CR density, the growth rate of waves $\Gamma_{\rm CR}W$ can be written following \cite{Skilling:1975p2176}:
\be
\Gamma_{\rm cr}(k)=\frac{16 \pi^{2}}{3} \frac{v_{\rm A}}{k\,W(k) B_{0}^{2}} \sum_{\alpha} \left[ p^{4} v(p) \frac{\partial f_{\alpha}}{\partial z}\right]_{p=Z_{\alpha} e B_{0}/kc} ,
\label{eq:gammacr}
\ee 
where $\alpha$ is the index labelling nuclei of different types. All nuclei, including all stable isotopes for a given value of charge, are included in the calculations. As discussed in  previous literature, it is very important to compute  the diffusion coefficient properly, and thus for a meaningful comparison with  the flux spectra and  secondary to primary ratios, notably B/C. The growth rate, written as in Eq.~(\ref{eq:gammacr}), refers to waves with wavenumber $k$ along the ordered magnetic field. It is basically impossible to generalize the growth rate to a more realistic field geometry by operating in the context of quasi-linear theory, therefore we  use here this expression but keep  its limitations in mind.

The diffusion coefficient relevant for a nucleus $\alpha$ can be written as
\be
D_{\alpha} (p) = \frac{1}{3} \frac{p\,c}{Z_\alpha eB_{0}} v(p) \left[ \frac{1}{k\ W(k)} \right]_{k=Z_{\alpha} e B_{0}/pc},
\label{eq:diff}
\ee
where $W(k)$ is the power spectrum of waves at the resonant wavenumber $k=Z_{\alpha} e B_{0}/p\,c$,  Z being the nuclear electric charge. The nonlinearity of the problem is evident here. The diffusion coefficient for each nuclear species depends on all other nuclei through the wave power $W(k)$, but the spectra are in turn determined by the relevant diffusion coefficient. 

In general, if a background of pre-existing waves is present in the system and evolving under the action of a cascade in $k$-space, the wave spectrum $W(k)$ satisfied an equation of the following type:
\be
\frac{\partial}{\partial k}\left[ D_{kk} \frac{\partial W}{\partial k}\right] + \Gamma_{\rm CR}W = q_{W}(k), 
\label{eq:cascade}
\ee
where $q_{W}(k)$ is the injection term of waves with wavenumber $k$  and $D_{kk}=C_{K} v_{A} k^{7/2} W(k)^{1/2}$ is the diffusion coefficient of waves in $k$ space for a Kolmogorov phenomenology and $C_{K}\approx 0.052$ is a numerical coefficient. Diffusion in $k$ space describes the cascading and in the absence of other phenomena (such as self-generation), it leads to the formation of a Kolmogorov spectrum. Interestingly, the quantity $D_{kk}/k^{2}$, which represents the time scale for cascading from a given wavenumber $k$, has the same functional form and the same numerical value as the rate of NLLD  usually quoted in the literature \cite{2003A&A...403....1P}. Hence, maybe improperly, we will refer to this process as NLLD below.  For the sake of simplicity we can limit our attention to the case in which pre-existing waves are only injected on a scale $l_{c}\sim 50-100$ pc (perhaps by supernova explosions). This means that $q_{W}(k)\propto \delta (k-1/l_{c})$. The level of pre-existing turbulence can be normalized to the total power $\eta_B=\delta B^{2}/B_0^2 = \int dk W(k)$. Strictly speaking the wavenumber that appears in this formalism is the one in the direction parallel to that of the ordered magnetic field. 

The transport of CRs in a background of both self-generated and pre-existing waves in the Galaxy was first investigated by \cite{2012PhRvL.109f1101B} and \cite{2013JCAP...07..001A}, where the main implications were discussed. In \cite{2015A&A...583A..95A} the authors calculated the spectrum of protons under the action of both the self-generated and pre-existing turbulence, and compared the results with the PAMELA data available at the time \cite{2011Sci...332...69A}. In \cite{2013JCAP...07..001A}, the study was extended to other nuclei and the B/C ratio was also computed. The presence of self-generated waves and pre-existing waves naturally leads to a spectral break in the spectra of nuclei at rigidity $\sim 100-1000$ GV. The calculations were later applied to the AMS-02 data \cite{2015PhRvL.115u1101A} and Voyager \cite{voyager} by \cite{2015A&A...583A..95A}, where the resulting B/C ratio was also compared withe preliminary AMS-02 data. 

   \begin{figure}
   \centering 
   \includegraphics[width=8.cm]{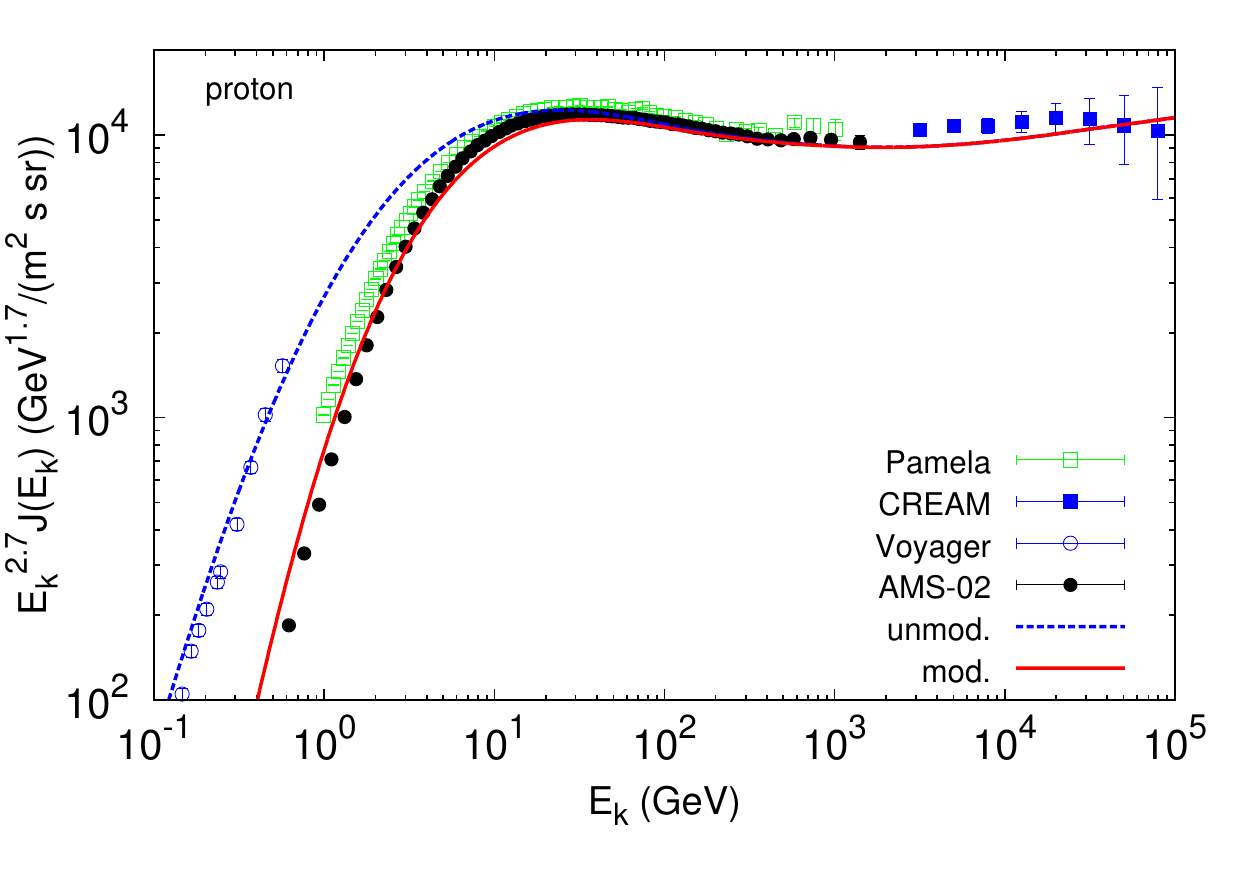}
   \caption{Spectrum of protons measured by Voyager (blue empty circles), AMS-02 (black filled circles) \cite[]{2015PhRvL.115u1101A}, PAMELA (green empty squares) \cite{2011Sci...332...69A} and CREAM (blue filled squares) \cite{cream}, compared with the prediction of our calculations (lines). The solid line is the flux at the Earth after the correction due to solar modulation, while the dashed line is the spectrum in the ISM. 
}
              \label{fig:spectrap}%
    \end{figure}

The spectrum of protons as calculated in \cite{2015A&A...583A..95A} is shown in Fig. \ref{fig:spectrap}. The solid lines indicate the spectra at the Earth, namely after solar modulation modelled using the force-free approximation \cite{gleeson68}, while the dashed lines are the spectra in the ISM. The data points are the spectra measured by the Voyager (empty circles) \cite{voyager}, AMS-02 (filled circles) \cite{2015PhRvL.115u1101A}, PAMELA (empty squares) \cite[]{2011Sci...332...69A}, and CREAM (filled squares) \cite[]{cream}. Fig. \ref{fig:spectrap} shows several interesting aspects: 1) the spectrum of protons (the same is seen in the He spectrum) shows a pronounced change of slope at few hundred GeV/n, where self-generation of waves becomes less important than pre-existing turbulence (in fact, the change of slope takes place in rigidity). 2) The spectra calculated to optimize the fit to the AMS-02 and PAMELA data is in excellent agreement with the Voyager data (see dashed lines). This is not trivial: in this model, at sufficiently low energies (below $\sim 10$ GeV/n), particle transport is dominated by advection (at the Alf\'en speed) with self-generated waves, rather than diffusion. This reflects into a weak energy dependence of the propagated spectra, which  is exactly what Voyager measured (see also \cite{potgieter2013}). 

The possibility that CRs may appreciably contribute to generating the waves responsible for their own scattering is very fascinating and pregnant with implications. Aside from the ones discussed above, one should recall a few other considerations that may be easily understood physically: first, wherever there are more sources of CRs, one should expect a higher gradient, hence more waves, which in turn result in a higher CR density. The connection between this simple line of thought and the problem of the CR gradient is intuitively clear and has been discussed recently by \cite{Recchia-grad} (see also contribution by G. Morlino, this volume). Second, it can be shown in a rather general way, that self-generation of waves automatically leads to a diffusion coefficient that at rigidities below $\sim 10$ GV becomes roughly energy independent to a value $D\sim 2 v_{A}H$, where $v_{A}$ is the Alfven speed and $H$ the size of the halo. The assumption that is usually made in transport calculations that the diffusion coefficient is constant in rigidity at low rigidity might in fact have a physical explanation in this non-linear effect. Even numerically, taking $v_{A}\sim 10$ km/s and $H\sim 3$ kpc, one would get $D\sim 2\times 10^{28}~\rm cm^{2}s^{-1}$, close to the value that is typically assumed in the literature to fit data. 

\section{CR induced Galactic winds}
\label{sec:Gal-wind}

Galactic winds may affect star formation, through the regulation of the amount of gas available \citep{2007MNRAS.377...41C,2013MNRAS.428..129S} and inject hot gas in the galactic halo. In our Galaxy there are indications of the existence of such hot gas from X-ray observations \cite{1994Natur.371..774B,1999A&A...347..650B}. Most important, a Galactic wind may affect the propagation of CRs \cite{Ptuskin:1997A&A...321..434P,Recchia-wind}, and in turn CRs may contribute to launching such winds \cite{1975ApJ...196..107I,1991A&A...245...79B,2008ApJ...674..258E}. The very fact that CRs have to leave the Galaxy in order to reach some stationary condition implies that a gradient in CR density (and pressure) must exist between the disc and the escape surface. The $-\nabla P_{CR}$ force acts in the direction pointing away from the Galactic disc, while the gravitational force exerted by dark matter, baryons and stars in the Galaxy, pulls material downward. If the CR induced force prevails on the gravitational pull close to the disc, then a Galactic wind may be launched. This is however only a necessary (but not sufficient) condition for launching a wind, in that material can be lifted up and fall down in what is known as Galactic fountains. A wind is launched only if the gas can be accelerated to supersonic speeds and its properties connected smoothly with the boundary conditions at infinity. 

The dynamics of CR induced winds, where CR propagation occurs because of their interaction with self-generated waves, is extremely complex and in the past some useful assumptions were made to simplify the problem. For instance, in the pioneering work of \cite{1991A&A...245...79B}, the authors assumed that the diffusivity of CRs is vanishingly small, so as to transform the transport equation of CRs in a fluid equation. Within this assumption, the so-called wind solutions were found but no information on the spectrum of CRs was retained. Later, \cite{Ptuskin:1997A&A...321..434P} proposed a simplified way to keep information on the CR spectrum and CR spatial distribution, and included Galactic rotation. Recently, \cite{Recchia-wind} proposed an iterative semi-analytical method to determine the wind structure, the CR spectrum and spatial distribution and to determine whether a CR induced wind may be launched in the first place, depending on the conditions at the base of the wind (Galactic rotation was not included in such calculation). 

A physical understanding of the way CR induced winds work was proposed by \cite{Ptuskin:1997A&A...321..434P}, and although their conclusions are usually not confirmed quantitatively by the calculations presented in \cite{Recchia-wind}, the physical intuition turns out to be very useful. 

The basic assumption of \cite{Ptuskin:1997A&A...321..434P} is that near the disc the advection velocity (dominated by the Alfv\'en speed) scales approximately linearly with $z$, $v_{A}\sim \eta z$. Now, it is easy to imagine that while the advection velocity increases with $z$, it reaches a critical distance, $s_{*}$, for which advection dominates upon diffusion. This happens when
\begin{equation}
\frac{s_{*}^{2}}{D(p)}\approx \frac{s_{*}}{v_{A}(s_{*})} \, \Rightarrow \, s_{*}(p) \propto D(p)^{1/2},
\end{equation}
where we used the assumption of linear relation $v_{A}\sim \eta z$ and that the diffusion coefficient has small spatial variation, which turns out to be true in self-generated scenarios. Now, when diffusion dominates, namely when $z\lesssim s_{*}(p)$, one can neglect the advection terms and make the approximate statement (as in the standard diffusion model), that $D(p)\frac{\partial f}{\partial z}|_{z=0}\approx -Q_{0}(p)/2 \propto p^{-\gamma}$. The quantity $s_{*}(p)$, that depends naturally on particle momentum, plays the role of the size of the diffusion volume and one can show that, similar to a leaky box-like model, the equilibrium spectrum in the disc is 
\begin{equation}
f(p)\sim \frac{Q_{0}(p)}{s_{*}(p)}\frac{s_{*}^{2}}{D(p)} \sim Q_{0}(p)\, D(p)^{-1/2}.
\end{equation}
In other words, $s_{*}(p)$ playes the role of $H$ in the standard model of CR transport. There are a few very important implications of this line of thought: 1) the energy dependence of the spectrum of CRs in the Galactic disc is not simply proportional to $Q_{0}(p)/D(p)$, as is usually assumed, but rather to $Q_{0}(p)\, D(p)^{-1/2}$, where however $D(p)$ is in turn a function of the CR spectrum. Using the expressions for the growth of Alfv\'en waves, Eq. \ref{eq:gammacr}, and their rate of damping, one can demonstrate that for $Q(p)\propto p^{-\gamma}$, $D(p)\propto p^{2\gamma-7}$, so that injection spectrum and observed spectrum are directly connected, at least for the energies where diffusion dominates. These expressions hold as long as NLLD is the main channel of wave damping. 2) The size H of the halo, where free escape should occur, loses its meaning: this is a positive outcome of these models since the boundary condition typically imposed at $|z|=H$ is somewhat artificial, and shows how weak the predictions of the so-called standard model of CR transport may be. The role of $H$ is played by the physical quantity $s_{*}(p)$ which however is an output of the problem (not imposed by hand) and depends on the particles' momentum. 

\begin{figure}
	\includegraphics[width=\columnwidth]{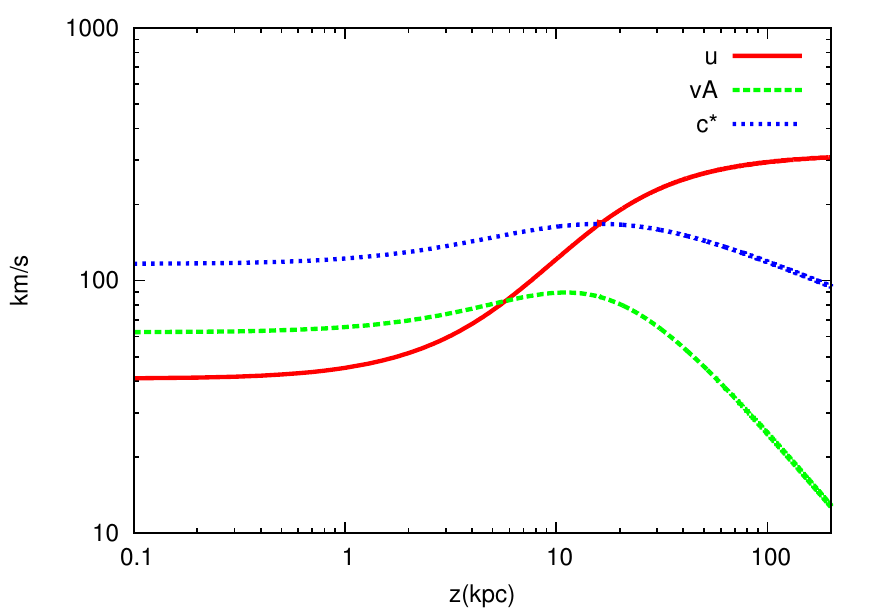}
    \caption{Wind velocity (red solid line), Alfv\'en velocity (green dashed line) and sound speed (blue dotted line) as a function of the height $z$ above the disc, for a reference case.}
    \label{fig:velocity}
\end{figure}

As pointed out above, the simple recipe proposed by \cite{Ptuskin:1997A&A...321..434P} does not apply to realistic situations because it is based on the assumption that at the base of the wind the advection velocity vanishes. In fact, the Alfv\'en speed and the wind velocity have a finite value at the base of the wind, which implies that at low energies the transport of CRs is dominated by advection. This can be seen in Fig. \ref{fig:velocity}, where we show the wind velocity (red solid line), Alfv\'en velocity (green dashed line) and sound speed (blue dotted line) as a function of the height $z$ above the disc for a typical wind case. 

\begin{figure}
	\includegraphics[width=\columnwidth]{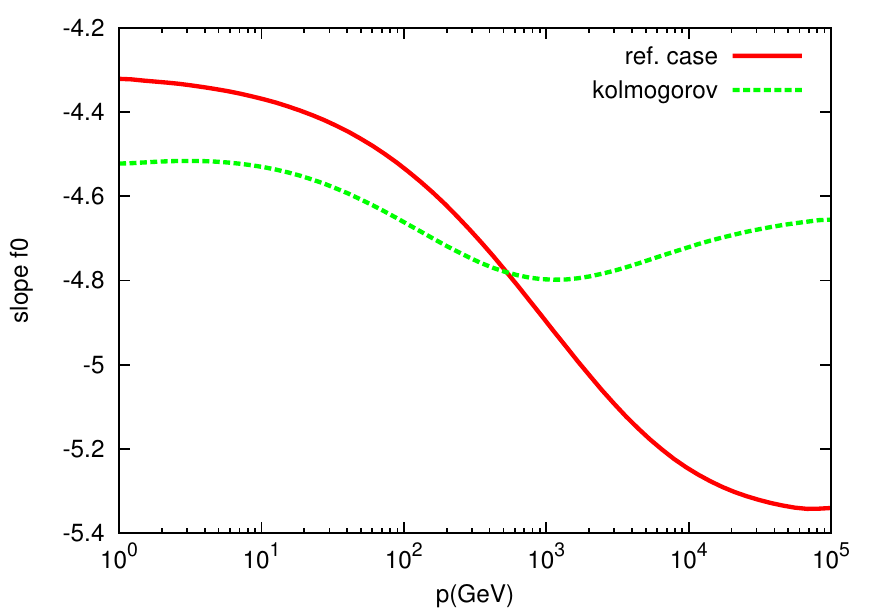}
    \caption{Slope of the CR spectrum in the disc of the Galaxy at the position of the Sun for the same case as in Fig. \ref{fig:velocity} (red solid line) and for the case in which the wind is launched at $z_{0}=1$ kpc from the disc and there is a near-disc region where the diffusion coefficient is assigned and advection is absent (green dashed line).}
    \label{fig:f0}
\end{figure}

This has important implications for the spectrum of CRs as observed at a location compatible with that of the Sun: as shown in \cite{Recchia-wind}, most wind solutions that may be found do not lead to CR spectra that are similar to the observed one, although they do correspond to dynamically feasible solutions with a substantial mass loss (of order $\sim 0.5~M_{\odot}~yr^{-1}$ integrated on the Galactic disc). For instance the slope of the spectrum of CRs at the Sun as calculated for the same wind model corresponding to Fig. \ref{fig:velocity} is shown as a (red) solid line in Fig. \ref{fig:f0}, clearly at odds with observations (even qualitatively). 

A rather generic problem of all transport models (including wind models) is that they are mostly based upon resonant interactions of charged particles with Alfv\'en waves. However, Alfv\'en waves are expected to be severely damped due to ion-neutral damping (a process physically due to charge exchange between neutral H atoms and protons). This damping is so severe that within $\sim 1$ kpc from the Galactic disc, where most neutral hydrogen is located, there can hardly be any Alfv\'en waves. For this reason, in original approaches to the wind problem (e.g. \cite{1991A&A...245...79B}) the effective CR plasma coupling was guaranteed by assuming that the wind be launched at $z_{0}\simeq 1$ kpc. This near-disc region is also the one that is most perturbed by the direct action of repeated supernova explosions: for this reason, one could speculate that this region may host a local, non self-generated, turbulence directly associated to hydrodynamical turbulence. In \cite{Recchia-wind}, the implications of this assumption were investigated in the case that the near-disc region is characterized by a Kolmogorov-like diffusion coefficient. In this scenario the base of the wind was assumed to be at $z_{0}=1$ kpc. In passing, we notice that the importance of this near-disc region was already recognized in \cite{1993A&A...269...54B}. 

The assumption of pre-existing (namely not self-generated) turbulence in the near disc region allows one to have spectra that qualitatively resemble the spectrum of CRs observed at the Earth, as shown by \cite{Recchia-wind}. This means that the low energy spectrum (below $\sim 10$ GeV) is dominated by advection with waves and is close to the injection spectrum; at energies above $10$ GeV the spectrum is fully determined by self-generated waves and has a slope close to the observed one; at energies above $\sim$ TeV, the spectrum hardens because this part of the spectrum is dominated by diffusion in the near disc region. This spectral shape, shown as a (green) dashed line in Fig. \ref{fig:f0}, is qualitatively similar to the one recently measured for protons and helium nuclei by PAMELA and AMS-02.  

It is important to keep in mind that the launching of a CR induced wind actually results from the interplay between CRs and gravity, so that the characteristics of the wind and even the possibility to launch a wind (rather than a time dependent outflow that may fall back onto the Galactic disc) depend on the gravitational potential, especially the one contributed by dark matter. A careful study of the dependence of the wind characteristics on the gravitational potential and the implications of CR induced winds in terms on mass loss from the Galaxy is being carried out (Recchia, Morlino \& Blasi, in preparation).

\section{Grammage around sources}
\label{sec:grammage-source}

As discussed above, the possibility that CRs may affect the environment in which they propagate, either through their dynamical action or through generation of plasma waves that in turn can affect their transport depends on the local CR gradients. In the Galaxy the relevant gradient is of order $\sim f_{0}/H$ where $f_{0}$ is the CR density in the disc, but close to the source of CRs (for instance a SNR), the CR density and the gradients can be much larger than this reference value and it is natural to expect that the non-linear effects discussed above become more severe. This phenomenon has been recently investigated by \cite{plesser,malkov,marta,nava1} though under somewhat different assumptions. In \cite{plesser} a semi-analytical self-similar solution of the transport equation was found in the assumption that NLLD was the only damping process to limit the growth of the waves. A self-similar solution was also found in \cite{malkov} in the assumption that ion-neutral damping (IND) limits the growth of self-generated waves. On the other hand \cite{marta} and \cite{nava1} proposed a numerical solution of the transport equation close to a CR source in the presence of both NLLD and IND. The paper of \cite{nava1} focuses on the implications of the self-generation in terms of confinement time close to the source that in turn may enhance the gamma ray emission due to CR interactions with a molecular cloud in the near source region. The paper by \cite{marta} emphasizes another interesting aspect of the problem: most CR sources are located in the disc of the Galaxy, where the typical density is $\sim 1\rm cm^{-3}$ and the magnetic field is mostly ordered on scales of $\sim 100$ pc, possibly larger if $\delta B/B<1$ on that scale. This means that, if self-generation is effective, CRs on their way to escape the near source region may accumulate an appreciable grammage: more specifically, if $\tau_{s}\simeq L_{c}^{2}/D_{sg}$ is the time spent to leave the region of size $L_{c}$ around the source, under the action of the self-generated diffusion coefficient $D_{sg}$, then the accumulated grammage becomes comparable with the Galactic grammage when $\tau_{s}\sim \tau_{H} h/H$, which is much smaller than the typical confinement time inferred from B/C or from unstable isotopes. In other words, even an escape time from the near source region of order $h/H\sim 10^{-2}$ smaller than the CR escape time from the Galaxy would allow CRs to accumulate a grammage in the near source region comparable with the one usually inferred from B/C observations. As shown by \cite{marta}, in the absence of neutral hydrogen, responsible for IND, the self-generation of waves is approximately sufficient to warrant these conditions. The density of neutral hydrogen in the ISM depends on the phase of the gas: the densest gas (the one that would provide more grammage) is also the coldest and thereby the one with more neutrals. Hence the effect of neutrals is most likely not negligible. 

On the other hand, following \cite{jean}, one could argue that in the WIM most gas is ionized with density that can be as high as $\sim 0.45 \rm cm^{-3}$ but neutral gas is still present with density $\sim 0.05 \rm cm^{-3}$. Such parameters would still be more than sufficient to quench the growth of Alfv\'en waves due to IND on time scales of relevance for propagation of CRs in the Galaxy. In the region around a SNR, the level of IND inferred for these parameters is still important but not as much as for Galactic CR propagation. Moreover, as argued in Ref. \cite{ferriere}, the WIM, having temperature $\sim 8000$ K, is expected to be made of fully ionized hydrogen, while only helium would be partially ionized. This latter picture would have prominent consequences in terms of IND, in that this process is due to charge exchange between ions and the partially ionized (or neutral) component, but the cross section for charge exchange between H and He is about three orders of magnitude smaller \cite{aladdin} than for neutral and ionized H, so that the corresponding damping rate would be greatly diminished. Unfortunately, at present, there is no quantitative assessment of this phenomenon and we can only rely on a comparison between cross sections of charge exchange. On the other hand, it is also possible that a small fraction of neutral hydrogen is still present, in addition to neutral helium: following \cite{ferriere}, the density of neutral H is $\lesssim 6\times 10^{-2} n_{i}$, and for $n_{i}=0.45  \rm cm^{-3}$ this implies an upper limit to the neutral H density of $\sim 0.03 \rm cm^{-3}$. 

In order to account, to some extent, for the uncertainty in the role of IND around sources of CRs, \cite{marta} considered the following cases: (1) No neutrals and gas density $0.45~\rm cm^{-3}$ (this is the best case scenario in terms of importance of the near source grammage, with the underlying assumption that the neutral component is entirely in the form of He); (2) Neutral density $n_{n}=0.05~\rm cm^{-3}$ and ion density $n_{i}=0.45~\rm cm^{-3}$; (3) Density of ionized H $0.45~\rm cm^{-3}$ and density of neutral H $0.03~\rm cm^{-3}$; (4) rarefied totally ionized medium with density $n_{i}=0.01~\rm cm^{-3}$. 

\begin{figure}
\begin{center}
{\includegraphics[width=\linewidth]{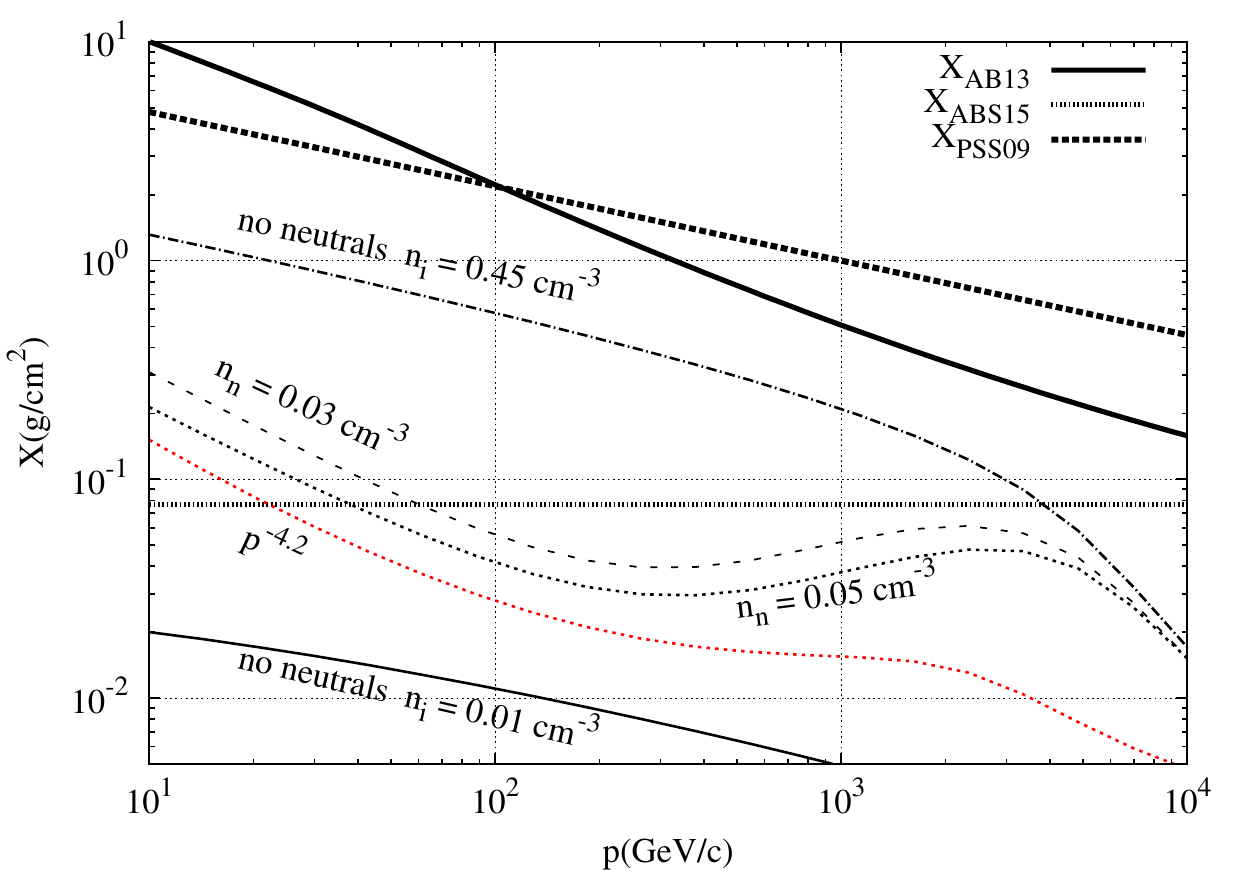}}
\caption{Grammage accumulated by CRs in the near-source region for $L_{c}=100$ pc in the three cases: (1) $n_{n}=0$, $n_i = 0.45 \mathrm{cm^{-3}}$; (2) $n_i = 0.45 \mathrm{cm^{-3}}$ and $n_n = 0.05 \mathrm{cm^{-3}}$; (3) $n_{n}=0$, $n_i = 0.01 \mathrm{cm^{-3}}$, as labelled. The thin dotted (red) line corresponds to case (2) but with slope of the injection spectrum 4.2. The thick dashed line (labelled as $X_{\rm PSS09}$) shows the grammage inferred from the measured B/C ratio \cite{fit}, while the thick solid line (labelled as $X_{\rm AB13}$) shows the results of the non-linear propagation of Ref. \cite{2013JCAP...07..001A}. The horizontal (thick dotted) line (labelled as $X_{\rm ABS15}$) is the source grammage, as estimated in Ref. \cite{2015A&A...583A..95A}.}
\label{fig:grammage}
\end{center}
\end{figure}

The grammage $X$ as a function of particle energy (or rigidity) is plotted in Fig.~\ref{fig:grammage}, for the case of a SNR with total kinetic energy $E_{SN}=10^{51}$ erg, field coherence length $L_c=100$ pc and CR acceleration efficiency $\xi_{CR}=20\%$. In terms of properties of the medium around the source, the four cases mentioned above are considered (see labels on the curves). 
The thick dashed line shows the grammage estimated from the measured B/C ratio, assuming standard CR propagation in the Galaxy with turbulence described {\it a la} Kolmogorov \cite{fit}. The thick solid curve represents the grammage as calculated in the model of non-linear CR propagation of Ref.~\cite{2013JCAP...07..001A} (see also \cite{2012PhRvL.109f1101B}), while the horizontal (thick dotted) line shows an estimate of the grammage traversed by CRs while still confined in the downstream of the SNR shock \cite{2015A&A...583A..95A}. In all cases the assumed injection was $\propto p^{-4}$, except for case (2) above, where the case of steeper injection was also considered (thin dotted (red) line).

Fig. \ref{fig:grammage} illustrates in a clear way how the phenomenon of self-confinement of CRs close to their sources can have potentially important implications on the interpretation of secondary-to-primary ratios in terms of grammage traversed in the Galaxy. The importance of these considerations depends in a critical way on the density of gas and the density of neutral hydrogen in the regions around sources, and on the size of the regions where the Galactic magnetic field can be considered as roughly ordered. 

\section{Discussion}
\label{sec:discuss}

The increasing confidence of the community in our modelling of CR propagation in the Galaxy is somewhat at odds with the equally increasing proposals of alternative ideas on CR transport, that attempt to question the very foundations of the so-called standard model. For instance, after the discovery of a rising positron fraction, \cite{waxman} proposed that the excess positrons do not result from the existence of unknown sources but rather from the poor knowledge of the secondary production of positrons in the ISM. The authors pointed out that a transport scenario could be arranged in which the positrons produced as a result of CR interactions in the ISM could fit observations. A similar proposal, though with several important differences in the details, was recently put forward in \cite{lipari}, where the author starts from the observation that the $\bar p/p$ ratio appears to be much harder than expected to conclude that positrons are also simply secondary products of CR interactions in the ISM in a model that is quite unlike the standard one. In this latter approach, leptons are required to not lose energy on the scale of the escape time, which implies that the escape time should be very short. As a consequence, the grammage measured through the B/C ratio is speculated to be accumulated near the sources, an idea previously put forward by \cite{2016ApJ...827..119C,2014ApJ...786..124C,2010PhRvD..82b3009C}. The scenario that arises from these models is usually not more economical in terms of assumptions, neither it is simpler in terms of underlying physics. Nevertheless these models need to be given proper consideration if there is a physical motivation to support them. One instance is the non-linear self-confinement of CRs near the sources, discussed in \S \ref{sec:grammage-source}, which suggests that under certain circumstances, at least a part of the grammage accumulated by CRs might be associated to propagation in the near source region. Of course these alternative scenarios need to be subject to the standard scrutiny of observation. For instance, the recent measurement of the B/C ratio by AMS-02 seems to question or at least put tension on models such as \cite{2016ApJ...827..119C}, although much caution needs to be used in that most expected differences between standard model and alternative models are confined to the rigidity region where systematic errors are the largest. 

Even putting aside radically new views of CR transport, it is a fact that the increasingly better data are showing interesting features (such as the breaks in the spectra of protons and helium nuclei) that require a physical explanation. These phenomena are best explained in terms of transport of CRs, either due to a spatially dependent diffusion coefficient \cite{2012ApJ...752L..13T} or due to the co-existence of self-generated and pre-existing turbulence \cite{2012PhRvL.109f1101B}. The possibility of self-confinement of CRs in the Galaxy or around their sources is a very fascinating concept that has numerous implications, all subject to observational testing, which makes this scenario very appealing. 

One of the most striking implications of this picture arises when one allows self-confined CRs to exert their dynamical action on the background plasma that they move in: in much the same way that the force of CRs on the plasma near a SNR shock leads to the formation of a CR precursor upstream of the shock (with all the consequences, such as concave spectra of accelerated particles), the force exerted by CRs on the ISM in the Galaxy can also move such plasma around. However, in the Galaxy this action is opposed by the gravitational pull, due to gas, stars and dark matter that try to pull the gas towards the center of the potential. As a result, an outflow may be launched if the $- \nabla P_{CR}$ force wins over gravity. The outflow may become supersonic and eventually lead to a CR driven wind, potentially able to lift off $\sim 0.5-1~\rm M_{\odot}$/yr from the Galactic disc. The implications of this phenomenon for star formation as well as for CR transport are remarkable, as discussed in \S \ref{sec:Gal-wind}. Interestingly the transport of CRs in a CR induced Galactic wind also solves one of the open problems of conventional approaches to CR propagation, namely the need to impose by hand the existence of a free escape boundary where particles are allowed to leak from the Galaxy. The predictions of these models in terms of CR spectrum at the Earth are however not easy to reconcile with observations (see discussion in \cite{Recchia-wind}).

Although not discussed in this paper, it is worth mentioning that non-linear self-confinement of CRs has also been discussed in the context of extragalactic sources of very high energy CRs \cite{martaUHECR}, where a current driven, non resonant instability is excited and may cause CRs with energy below $\sim 10^{7}-10^{8}$ GeV to be confined close to their sources for times exceeding the age of the universe, thereby causing the appearance of a {\it low-energy} cutoff in the spectra observed at the Earth. 

\section*{Acknowledgements} 
The author is very grateful to E. Amato for reading the manuscript and for continuous scientific collaboration. The author also expresses gratitude to the organizing committee of the workshop ``Cosmic Ray Origin - beyond the standard models'' for creating an exciting atmosphere in a charming setting.







\end{document}